\documentclass[runningheads]{llncs}
\usepackage{graphicx}
\usepackage{subfigure} 
\usepackage{algorithm}
\usepackage{amsmath}
\usepackage{algorithmic}
\usepackage{graphicx} 
\usepackage{booktabs}
\usepackage{amsfonts,amssymb}
\usepackage{booktabs}
\usepackage{multirow}
\usepackage{threeparttable}
 \usepackage{mathrsfs}

\begin{document}
%
% \title{Joint Few-Shot Learning with Side Information for Skin Disease Classification}
%\title{Few-shot Skin Disease Classification: On Alleviating the Incompatibility between Cross Entropy Loss and Episode Training}
\title{Alleviating the Incompatibility between Cross Entropy Loss and Episode Training for Few-shot Skin Disease Classification}
\titlerunning{Few-shot Skin Disease Classification}
\author{Wei Zhu\inst{1} \and
Haofu Liao\inst{1} \and
Wenbin Li\inst{2} \and
Weijian Li\inst{1} \and
Jiebo Luo\inst{1}}
\institute{University of Rochester \and
Nanjing University}
\maketitle              % typeset the header of the contribution
\begin{abstract}
Skin disease classification from images is crucial to dermatological diagnosis. However, identifying skin lesions involves a variety of aspects in terms of size, color, shape, and texture. To make matters worse, many categories only contain very few samples, posing great challenges to conventional machine learning algorithms and even human experts. Inspired by the recent success of Few-Shot Learning (FSL) in natural image classification, we propose to apply FSL to skin disease identification to address the extreme scarcity of training sample problem. However, directly applying FSL to this task does not work well in practice, and we find that the problem can be largely attributed to the incompatibility between Cross Entropy (CE) and episode training, which are both commonly used in FSL. Based on a detailed analysis, we propose the Query-Relative (QR) loss, which proves superior to CE under episode training and is closely related to recently proposed mutual information estimation. Moreover, we further strengthen the proposed QR loss with a novel adaptive hard margin strategy. Comprehensive experiments validate the effectiveness of the proposed FSL scheme and  the possibility to diagnosis rare skin disease with a few labeled samples. 

\end{abstract}

\section{Introduction}
As a key step in the dermatological diagnosis, skin disease classification is quite challenging due to the extremely scarce annotations for a large number of categories. Such complexity in skin disease taxonomy requires a great deal of expertise. In addition, the diagnosis is often subjective and inaccurate even by human experts, which necessitates the research for computer-aided diagnosis \cite{okuboyejo2013automating,sumithra2015segmentation}. Motivated by the unprecedented success of deep neural networks (DNNs), many researchers resort to deep learning technologies to handle this task \cite{esteva2017dermatologist,liao2016skin,liao2018deep}. For example, Esteva \textit{{et al.}} adopt GoogleNet Inception V3 \cite{szegedy2016rethinking} to train a large-scale skin disease classification network \cite{esteva2017dermatologist}. Liao \textit{et al.} jointly train skin lesion and body location classifiers using a multi-task network \cite{liao2018deep}. However, since DNN-based methods usually require a significant number of training samples for each category, categories with only a few number of samples are often discarded \cite{liao2016skin}. This reduces the applicability of DNN-based methods, especially for infrequent skin disease diagnosis. 

 Shi \textit{et al.} propose to adopt active learning to reduce the annotation cost \cite{DBLP:conf/miccai/ShiDX00H19}, but still need up to 50\% of labeled samples to train their model. Alternatively, Few-Shot Learning (FSL) is usually leveraged to address such tasks with only a few training samples \cite{snell2017prototypical,Sung_2018_CVPR,li2019revisiting,li2019distribution}. By assuming the availability of a large-scale auxiliary training set, one can learn generalized patterns and knowledge which facilitate the learning for unseen tasks. Formally, for each few-shot task, we are provided with a support set $S$, a query set $Q$, and an auxiliary set $A$, where the support set $S$ contains $C$ different categories and each category has $K$ training samples, \textit{i.e.}, $C$-way $K$-shot, and $Q$ contains unlabeled query data. Instead of conventional minibatch training, FSL is always trained with the episode training mechanism \cite{snell2017prototypical}. Basically, at each training iteration, we generate an episode by drawing samples from $C$ different categories of the auxiliary set $A$, with $K$ samples in each category as support samples $S_{train}$ and others as query samples $Q_{train}$. As a crucial step, we need to randomly shuffle the labels for all categories from episode to episode. Episode training mechanism benefits FSL in at least two aspects. First, it enables FSL to be trained under similar scenarios as testing tasks. Second, the labels are randomly shuffled during episode training, which enables the model to learn category-agnostic representation for a better generalization ability.   

Generally, FSL employs the Cross Entropy (CE) loss as an objective for classification. Although CE is useful for conventional classification, we find that it is somewhat incompatible with the episode mechanism. Well-designed FSL methods trained with CE even perform significantly worse than the baseline methods \cite{chen2019closerfewshot}. As we will see, CE classifies the query samples individually and relies highly on well-trained category-wise representation, a.k.a. proxies in proxy-based metric learning methods which share the similar formulation as CE \cite{movshovitz2017no,qian2019softtriple}. The proxy is an category-wise \textit{aggregation} of labeled support samples, e.g., the center used in Prototypical Network (PN). However, accurate proxies could only be obtained by a large-scale unified labeled dataset under the conventional minibatch training mechanism. This is hardily fulfilled under the episode training mechanism since we are only provided with a few training samples with randomly shuffled labels in each iteration. 
% For example, the proxies are often replaced by the mean of the few training samples of each episode, and are thus clearly highly biased and inaccurate.  
% We try to understand the problem form the aspect of proxy-based metric learning which shares similar formulation with CE for normalized vector \cite{qian2019softtriple}. 

To alleviate the problem, we propose a Query-Relative (QR) loss, which works much better with the episode training mechanism than CE for FSL. 

We highlight our main contributions as follows:
\vspace{-0.0em}
\begin{itemize}
    \item Upon an insightful analysis of the CE loss and episode mechanism, we propose a Query-Relative (QR) loss to better utilize the cross sample information and avoid possible sub-optimal aggregation of negative support samples, which significantly boosts the FSL performance;
    \item We develop an adaptive hard margin method for the QR loss to further penalize the categories with more error similarity connections;
    \item We evaluate our methods against a benchmark FSL suite \cite{chen2019closerfewshot}, and the experiments strongly validate our analysis and the proposed methods. 
    %%% what do you mean by a benchmark platform? it does not look medically related either
\end{itemize}
%The QR makes better use of the rich cross sample information within each episode and thus reduces the dependencies on accurate proxies.
%Moreover, we avoid manually designed support sample aggregation to some extent and allow the model to learn to extract information from support samples under the guidance of training objective. We further equip QR with an adaptive hard margin method, which enables the model to focus on the categories containing more errors.

\begin{figure}
\centering
\includegraphics[width=0.9\textwidth]{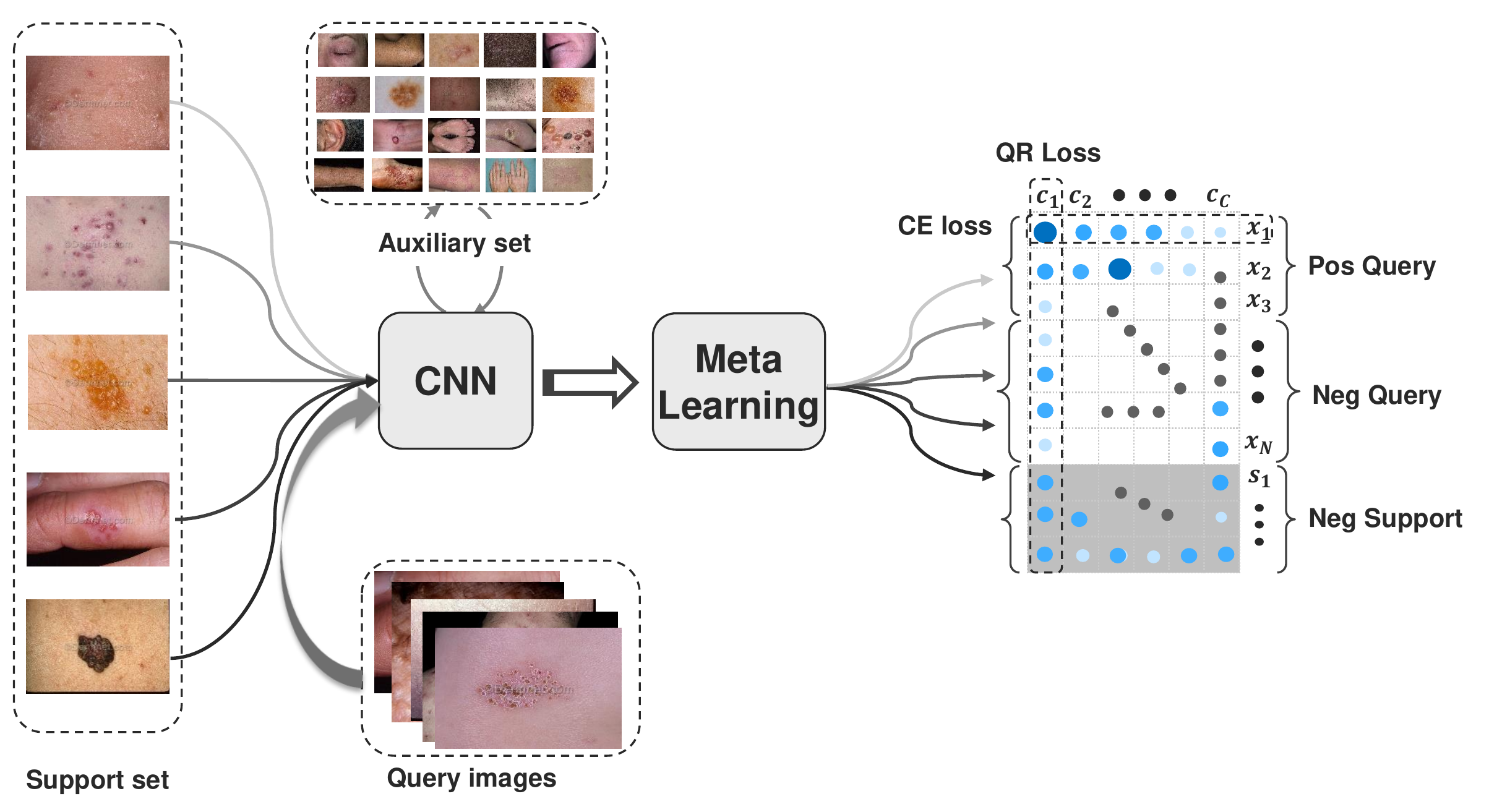}
\caption{Block diagram of few-shot learning-based skin disease classification and the difference between QR and CE loss. CE considers  queries individually, while QR takes the relation across samples into consideration. Moreover, CE aggregates the support samples into proxies with possible information loss, while QR allows the model to fully exploit the information of negative support samples guided by the training objective.}
\label{fig:dig}
\end{figure}
% \section{Related Work}

\section{Methodology}

\subsection{Discussions on FSL}
Cross Entropy (CE) loss is often jointly used with episode mechanism to solve the FSL tasks. It can be generally formulated as 
\begin{equation} \label{eq:ce}
    \mathcal{L}_{\text{CE}} = -\sum_i \log \frac{e^{s(c_{y_i}, x_i)}}{\sum_j e^{s(c_{j}, x_i)}}.
\end{equation}
Here, $\lbrace x_1, x_2, \dots, x_N \rbrace \in \mathbb{R}^{d \times N}$ are the query embeddings, and $\lbrace c_1, c_2, \dots, c_C \rbrace \in \mathbb{R}^{d \times C}$ are the representations for the support categories, where $N$, $C$, and $d$ denote the number of queries, support categories, and feature dimensions, respectively. $s(c_j,x_i)$ denotes the similarity between the support category proxy $c_j$ and query sample $x_i$. Different FSL methods have different formulations of similarity measurement $s(\cdot,\cdot)$ and category proxy $c$ aggregated by the support samples. For example, PN (Prototypical Network) uses the centers of the support samples from the $j$-th category as $c_j$ and the Euclidean distance as $s(\cdot,\cdot)$, Matching Net employs an FCE (Fully Context Embedding) layer to encode the support samples and chooses cosine similarity for $s(\cdot, \cdot)$, and MAML implements $s(\cdot,\cdot)$ as a Fully Connected (FC) layer where the $j$-th weight vector of the FC layer corresponds to $c_j$. To unify these methods, we normalize $x_i=\frac{x_i}{\Vert x_i \Vert_2}$ and $c_j=\frac{c_i}{\Vert c_i \Vert_2}$ which leads to better performance shown in the recent literature \cite{ye2020identifying}. Eq. (\ref{eq:ce}) can then be rewritten as  $\min - \sum_i \log \frac{e^{c_{y_i}^T x_i}}{\sum_j e^{c_{j}^T x_i}}$.

According to Eq. (\ref{eq:ce}), CE individually classifies the query samples and completely relies on the category-wise representation $c_j$ to train the model. For the conventional classification task trained with minibatch SGD, such a mechanism could prompt $c_j$ to learn high-level representative features of each category by exposing them to a large and balanced dataset. Unfortunately, this is not the case for FSL due to the episode training mechanism. Although episode training is important for FSL since it empowers FSL with the ability to learn generalized class agnostic representation and provides similar training scenarios as testing scenarios, it is also a double-edged sword: it makes $c_j$ inevitably biased and inaccurate. The reasons are two aspects: first, $c_j$ is learned from a few samples in each episode, e.g., 1 and 5 for $1$-shot and $5$-shot respectively, and it is difficult to learn to aggregate support samples to obtain $c_j$ without losing useful information with so few training samples; second, the labels are randomly shuffled for each episode which limits $c_j$ to be consistently trained across episodes. Therefore, $c_j$ cannot be fully relied on under the episode mechanism and training the model with CE loss will eventually degrade the performance for FSL. The sub-optimal performance has been observed and experimentally validated by several recent benchmark papers for natural images, where well-designed baselines could achieve similar and even better performance than CE-trained FSL counterparts \cite{chen2019closerfewshot,triantafillou2019meta}. Similar results are also disappointingly observed in the skin disease tasks according to our experiments in Section \ref{sec: experiment}. 

\subsection{Query-Relative Loss}
We alleviate the above problem from two aspects. First, instead of classifying the query separately, we unify all samples into a joint objective to allow them to mutually share information cross samples. Second, we avoid using negative category proxies which are aggregated by the negative support samples with a manually designed strategy (e.g., the center of support samples in PN), and the information of the support samples can then be largely preserved and extracted with the guidance of the training objective. To this end, we propose the Query-Relative (QR) loss as follows
\begin{equation} \label{eq:our}
     \mathcal{L}_{\text{QR}} = \sum_j \log (1+ \frac{1}{2|P_j|}\sum_{x_i^+\in P_j} e^{-s(c_{j},x_i^+)} + \frac{1}{2|N_j|} \sum_{x_i^-\in N_j} e^{s(c_{j}, x_i^-)}),
\end{equation}
where $P_j$ denotes the set of positive query samples that belong to the $j$-th category and $N_j$ denotes the set of \textit{negative query and support} samples that are not from the $j$-th category. $|\cdot|$ denotes the number of samples in the set.

We then present an analysis on how our objective improves CE from the two aforementioned aspects. First of all, Eq. (\ref{eq:our}) implicitly utilizes the cross sample information to re-weight each sample. Specifically, taking the derivation w.r.t. $s(c_j, x_p^+)$ and $s(c_j, x_n^-)$, we have 
\begin{equation} \label{eq:de}
\begin{split}
\biggr \vert \frac{\partial \mathcal{L}}{\partial s(c_j, x_p^+)}\biggr \vert = &\frac{\frac{1}{2|P_j|}e^{-s(c_{j},x_p^+)}}{1+ \frac{1}{2|P_j|}\sum_{x_i^+\in P_j} e^{-s(c_{j},x_i^+)} + \frac{1}{2|N_j|} \sum_{x_i^-\in N_j} e^{s(c_{j}, x_i^-)}} \\
% =&\frac{1}{|P_j|e^{s(c_{j},x_p^+)}+ \sum_{x_i^+\in P_j} e^{s(c_{j}, x_p^+)-s(c_{j},x_i^+)} + \frac{1}{2|N_j|} \sum_{x_i^-\in N_j} s(c_{j}, x_i^+)+e^{s(c_{j}, x_i^-)}} 
\end{split}
\end{equation}
\begin{equation}\label{eq:4}
    \biggr \vert \frac{\partial \mathcal{L}}{\partial s(c_j, x_n^-)}\biggr \vert = \frac{\frac{1}{2|N_j|}e^{s(c_{j},x_n^-)}}{1+ \frac{1}{2|P_j|}\sum_{x_i^+\in P_j} e^{-s(c_{j},x_i^+)} + \frac{1}{2|N_j|} \sum_{x_i^-\in N_j} e^{s(c_{j}, x_i^-)}}\,\\
\end{equation}
% For simplicity, we only give an discussion on Eq. (\ref{eq:de}), and reader can apply similar analysis . 
Here, we only focus on the absolute value of the gradient. According to Eq. (\ref{eq:de}), ${s(c_{j},x_p^+)}$ will induce a large gradient and will be punished if (i) ${s(c_{j},x_p^+)}$ is small; (ii) ${s(c_{j},x_p^+)}$ is smaller than ${s(c_{j},x_i^+)}$  where $x_i^+ \in P_j, \; p \neq i$; or (iii) $s(c_j, x_i^-)$ is small so that we could focus on intra-class relation. Moreover, a large $s(c_{j}, x_i^-)$ will provide  ${s(c_{j},x_i^+)}$ with tolerance to some extent, which allows our model to focus on reducing the large similarity of $s(c_{j}, x_i^-)$. Similar analysis can be performed with $s(c_{j},x_n^-)$ based on Eq. (\ref{eq:4}), and we omit the detail here. Therefore, in contrast to CE which deals with each sample separately, QR allows the query and support samples to share information across each other and category-wisely re-weights their importance. 

Second, note that the negative set $N_j$ of each category contains not only the negative query samples but also the support samples from other categories. This avoids the information loss caused by the possibly sub-optimal support sample aggregation and allows the model to learn to utilize the negative support samples directly by the objective.   

It turns out that the QR loss is closely related to Deep Mutual Information (MI) maximization recently proposed by \cite{hjelm2018learning}. Without loss of generality, following \cite{hjelm2018learning}, the JSD-based (Jensen-Shannon Divergence) MI estimator between $c_j$ and $x$ can be formulated as  
\begin{equation}\label{eq:mi}
\begin{split}
        \mathcal{L}_{\text{JSD MI}} &=  \max \frac{1}{|P_j|}\sum_{x_i^+\in P_j}-\log(1+e^{-s(c_{j},x_i^+)}) - \frac{1}{|N_j|} \sum_{x_i^-\in N_j} \log( 1+ e^{s(c_{j}, x_i^-)})\\
        &\geq \max -\log(1+\frac{1}{2|P_j|}\sum_{x_i^+\in P_j}e^{-s(c_{j},x_i^+)} + \frac{1}{2|N_j|} \sum_{x_i^-\in N_j} e^{s(c_{j}, x_i^-)})\\
        &= \mathcal{L}_{\text{QR}}
\end{split}
\end{equation}
Here we use the fact that $-\log(1+x)$ is convex and the Jensen's inequality. We can thus derive that the QR loss is actually a lower bound of the JSD MI. The reason why we do not directly optimize $\mathcal{L}_{\text{JSD MI}}$ is that the re-weighting mechanism of $\mathcal{L}_{\text{JSD MI}}$ does not take both $P_j$ and $N_j$ into consideration for each $s(c_j, x_i)$. We experimentally verify the superiority of our formulation in Sec. \ref{sec: experiment}. 
%$\mathcal{\hat{I}}({c_j,x})$

\subsection{Adaptive Hard Margin}
The adaptive hard margin is built upon the fact that the cosine similarity between uniformly distributed normalized samples approaches $\mathcal{N}(0, \frac{1}{2d})$ \cite{wu2017sampling} and is thus likely to be zero. Therefore, $s(c_{j},x_i^+)$ should be at least larger than $E_{j}^-$ and $s(c_{j},x_i^-)$ should be at least smaller than $E_{j}^+$, where $E_{j}^+$ and $E_{j}^-$ denote the average of $s(c_{j},x_i^+)$ with $s(c_{j},x_i^+)<0$ and average of $s(c_{j},x_i^-)$ with $s(c_{j},x_i^+)>0$, respectively. Based on this observation, we propose a QR loss with online Adaptive Hard Margin which can be written as
\begin{equation}\label{eq:hard}
 \mathcal{L}_{\text{QR+margin}} = \sum_j \log (1+ \frac{1}{2|P_j|}\sum_{x_i^+\in P_j} e^{-s(c_{j},x_i^+) + E_{j}^-} + \frac{1}{2|N_j|} \sum_{x_i^-\in N_j} e^{s(c_{j}, x_i^-)-E_{j}^+})\,.
\end{equation}
Basically, Eq. (\ref{eq:hard}) imposes extra punishment on categories with more positive samples whose similarities are smaller than random or negative samples, and negative samples whose similarities are larger than random or positive samples.  

\section{Experiments} \label{sec: experiment}

\subsection{Datasets}
We collect the dermatology images from the Dermnet atlas website \footnote{{www.dermnet.com}}. To perform few-shot learning, we discard categories with less than $10$ samples, which are required for the 5-way 5-shot setting. Finally, we obtain $20,230$ images in total belonging to $334$ different categories. The largest category ``seborrheic keratoses ruff" contains $516$ images and the smallest categories contain 10 samples. Detailed statistics of the data can be found in the supplemental material. The data is manually split into $186$ categories for training, $74$ for validation, and $74$ for testing, respectively. Moreover, to better simulate the scenario of few-shot learning, we deliberately choose categories with more than $120$ samples ($38$ categories in total) as the  training data.

\subsection{Benchmark Methods and Experimental Settings}
We benchmark the dataset on an FSL suite proposed by \cite{chen2019closerfewshot}. The suite contains 2 strong baseline methods (denoted as baseline and baseline++ following \cite{chen2019closerfewshot}) and 4 FSL methods including Relation Net\cite{Sung_2018_CVPR}, Model-Agnostic Meta-Learning (MAML) \cite{finn2017model}, Matching Net (MN)\cite{vinyals2016matching}, and Prototypical Net (PN)\cite{snell2017prototypical}. The baseline methods are carefully designed and outperform FSL methods in some cases. We refer readers to \cite{chen2019closerfewshot} for details. The four FSL methods are regarded as the state-of-the-art FSL baselines in recent benchmark literature \cite{triantafillou2019meta,chen2019closerfewshot}, and we train them with CE as our baselines except for the Relation Net, which is trained with Mean Square Error (MSE) Loss following the original paper. We apply the proposed QR loss to MN and PN since these two methods have proven to have superior and stable performance in natural image classification \cite{chen2019closerfewshot}. The model trained with JSD-based MI maximization Eq. (\ref{eq:mi}) is denoted as JSD MI, and models trained with the proposed QR loss Eq. (\ref{eq:our}) and QR loss with adaptive hard margin Eq. (\ref{eq:hard}) are denoted as QR and QR+M, respectively. 

For the network structure, we follow the commonly adopted FSL settings \cite{Sung_2018_CVPR,chen2019closerfewshot}. The feature embedding network used in this paper is a convolutional neural network which has four convolutional blocks with each block containing a sequence of a convolutional layer with 64 filters of size $3\times 3$, a batch normalization layer, a $2\times 2$ max-pooling layer and a Leaky ReLU layer.
For the experimental settings, the episodic training mechanism is applied to all FSL models, and $60,000$ episodes are constructed in total during training for all methods. For validation and testing, 600 episodes are randomly constructed from the validation and test set, respectively. We conduct 5-way 1-shot and 5-way 5-shot classification tasks on the collected Dermnet dataset, and 5 query samples are provided for each category within each episode for either training and testing. For optimization, we adopt the Adam algorithm with a learning rate of $0.001$. Experiments are run five times and we report the performance on test set corresponding to the best validation results. The average Accuracy, Precision, and F1 score with 95\% confidence interval are reported.

\begin{table}%[!htbp]
\vspace{-3mm}
\centering
\begin{tabular}{@{}lcccccc@{}}
\toprule
\multirow{2}{*}{Methods} & \multicolumn{3}{c}{5-way 1-shot}      & \multicolumn{3}{c}{5-way 5-shot}     \\ \cmidrule(l){2-7} 
                         & ACC\%    & Precision\%  & F1\%         & ACC\%   & Precision\%  & F1\%         \\ \midrule
Baseline                 & 39.89 \tiny{$\pm 0.89$}  & 40.57\tiny{$\pm1.12$} & 37.16\tiny{$\pm0.91$} & 59.87\tiny{$\pm0.94$} & 62.37 \tiny{$\pm1.10$} & 58.19 \tiny{$\pm1.01$} \\ 
Baseline++               & 42.47 \tiny{$\pm0.94$} & 43.70\tiny{$\pm1.11$} & 40.34\tiny{$\pm0.93$} & 63.37\tiny{$\pm0.95$} & 65.80 \tiny{$\pm1.07$} & 61.75 \tiny{$\pm1.01$} \\ \midrule 
MAML                     & 45.95 \tiny{$\pm1.06$}  & 44.82\tiny{$\pm1.29$} &42.18 \tiny{$\pm1.08$} & 66.93 \tiny{$\pm0.96$} & 69.24 \tiny{$\pm1.11$} & 64.92 \tiny{$\pm1.05$} \\ 
Relation Net             & 45.50 \tiny{$\pm1.07$}  & 46.36\tiny{$\pm1.18$} & 44.00\tiny{$\pm 1.07$} & 62.53 \tiny{$\pm1.02$} & 64.90 \tiny{$\pm1.11$} & 62.26 \tiny{$\pm1.05$} \\ \midrule
%DN4             & 52.28 \tiny{$\pm1.09$}  & 53.87\tiny{$\pm1.26$} & 49.77\tiny{$\pm 1.13$} & 75.10 \tiny{$\pm0.89$} & 77.89 \tiny{$\pm0.92$} & 73.97 \tiny{$\pm 0.95$} \\
%CovaM Net             & 51.09 \tiny{$\pm1.05$}  & 52.10\tiny{$\pm1.27$} & 47.95\tiny{$\pm 1.10$} & 65.75 \tiny{$\pm1.02$} & 68.88 \tiny{$\pm1.08$} & 63.98 \tiny{$\pm1.02$} \\ \midrule
MN         & 44.59\tiny{$\pm0.97$}        & 44.96\tiny{$\pm1.19$}      & 41.52\tiny{$\pm1.00$}      &  61.21\tiny{$\pm0.90$}          &63.15 \tiny{$\pm1.13$}            & 58.29\tiny{$\pm0.99$}           \\ 
MN+JSD MI     & 43.28 \tiny{$\pm1.04$}  & 43.26\tiny{$\pm1.25$} & 40.00\tiny{$\pm 1.05$} & 58.99 \tiny{$\pm 0.94$}  & 60.23\tiny{$\pm1.20$} & 55.78\tiny{$\pm 1.02$}  \\ \cmidrule(l){1-7}
MN+QR     & 48.01 \tiny{$\pm1.09$}  & 48.87\tiny{$\pm1.13$} & 44.30\tiny{$\pm 1.13$} & 67.09\tiny{$\pm 0.97$} & 69.18 \tiny{$\pm1.16$} & 64.53 \tiny{$\pm1.08$} \\
MN+QR+M  & 49.29 \tiny{$\pm1.31$}  & 49.95\tiny{$\pm1.05$} & 45.64\tiny{$\pm 1.09$}  & 66.83\tiny{$\pm 0.95$} & 69.10 \tiny{$\pm1.16$} & 64.25 \tiny{$\pm1.05$} \\ 
MN+QR*    & 48.66 \tiny{$\pm1.07$}  & 48.86\tiny{$\pm1.30$} & 44.98\tiny{$\pm 1.11$} & - & - & - \\
MN+QR+M*        & 49.76 \tiny{$\pm1.07$}  & 49.52\tiny{$\pm1.32$} & 46.01\tiny{$\pm 1.13$} & - &- & - \\ \midrule

PN                       & 46.77 \tiny{$\pm1.04$}  & 46.82\tiny{$\pm1.06$} & 43.58\tiny{$\pm 1.07$} & 62.06\tiny{$\pm 1.02$} & 63.39 \tiny{$\pm1.22$} & 59.50 \tiny{$\pm1.10$} \\ 
PN+JSD MI                & 47.55 \tiny{$\pm1.00$}  & 47.90\tiny{$\pm1.25$} & 44.33\tiny{$\pm 1.05$} & 61.15\tiny{$\pm 0.94$} & 61.74\tiny{$\pm 1.16$} & 58.34\tiny{$\pm 1.02$} \\ \cmidrule(l){1-7}
PN+QR                  & 49.85 \tiny{$\pm1.11$}  & 49.53\tiny{$\pm1.32$} & 46.34\tiny{$\pm 1.14$} & 70.38 \tiny{$\pm0.96$} & 72.13 \tiny{$\pm1.08$} & 68.50 \tiny{$\pm1.05$} \\ 
PN+QR+M           & \textbf{52.41 \tiny{$\pm1.09$} } & \textbf{53.21\tiny{$\pm1.27$}} & \textbf{49.52 \tiny{$\pm1.12$} }& \textbf{71.99\tiny{$\pm 0.87$}} & \textbf{74.23 \tiny{$\pm0.98$}} & \textbf{70.30 \tiny{$\pm0.94$}} \\ 
PN+QR*                 & 50.62 \tiny{$\pm1.10$}  & 50.83\tiny{$\pm1.32$} & 47.16 \tiny{$\pm1.13$} &   -         &  -          &  -          \\
PN+QR+M*          & \underline{53.30 \tiny{$\pm1.11$}}  & \underline{53.69\tiny{$\pm1.35$}} & \underline{50.45 \tiny{$\pm1.17$}} &   -         &    -        &    -        \\ \bottomrule
\end{tabular}
\caption{Experimental results on the Derment skin disease classification dataset. * denotes that the model is trained with 9 query samples per episode. - denotes that the setting is not applicable. M denotes our methods with an adaptive hard margin.}
\vspace{-8mm}
\label{tab:results}
\end{table}

\subsection{Result Analysis}
The experimental results are reported in Table \ref{tab:results}, and we draw several interesting points from the results as follows. First of all, the baseline methods with minibatch training and CE loss perform reasonably well in practice. The FSL methods trained with CE loss have comparable or slightly better performance. In contrast, FSL methods trained with the proposed QR loss significantly outperform the baseline methods and the FSL methods with CE. For Matching Net, our QR loss achieves 3.42 \% and 5.88 \% improvements compared with the CE loss in terms of accuracy for 5-way 1-shot and 5-way 5-shot tasks. Significant improvements are also observed for PN, and our QR loss outperforms CE 3.08 \% and 8.32 \% for 5-way 1-shot and 5-way 5-shot, respectively. The improvements are obtained by fully utilizing the cross-sample information and avoiding the information loss caused by manually designed support sample aggregation during training. Second, we compare the QR loss with JSD MI. Although the formulations are similar, QR is significantly better than JSD MI. The reason should be attributed to the fact that JSD MI does not mutually utilize the information in $P_j$ and $N_j$. Third, the adaptive hard margin consistently boosts the performance of the models trained by QR. For example, the adaptive hard margin improves PN trained with QR 2.56 \% and 1.61 \% for 5-way 1-shot and 5-way 5-shot, respectively. Finally, our method could be further boosted by increasing the number of queries for both training and testing. Overall, our proposed FSL methods classify skin disease with only a few available training samples and makes it possible to diagnose rare diseases using modern neural networks.   

\subsection{Influence of the number of shots and ways}
For simplicity, we only conduct experiments on PN with various ways and shots and report the accuracy. As shown in Tables \ref{tab:multishot} and \ref{tab:multiway}, QR has clear advantages over CE when more samples are available per episode, suggesting that QR can better utilize the cross sample information.
\begin{table}[!htb]
    \begin{minipage}{.5\linewidth}
      \centering
        \begin{tabular}{@{}lccccc@{}}
        \toprule
        {\# shots} & 1 & 2     & 3     & 4 & 5 \\ \midrule
        CE       &46.77   & 54.04 & 57.15 &59.65   &62.06   \\ \midrule
        QR       &49.85   & 62.37 & 66.87 &68.95   &70.38   \\ \bottomrule
        \end{tabular}
        \caption{5-way different-shot. (ACC\%)}
        \label{tab:multishot}
    \end{minipage}%
    \begin{minipage}{.5\linewidth}
      \centering
        \begin{tabular}{@{}lccccc@{}}
        \toprule
        {\# ways}  & 2  &3   & 5      & 10 &20\\ \midrule
        CE       &69.80    & 59.42 & 46.77    &35.78 &24.30   \\ \midrule
        QR       &72.02    & 61.57 & 49.85    &40.37  &31.31 \\ \bottomrule
        \end{tabular}
        \caption{Different-way 1-shot. (ACC\%)}
        \label{tab:multiway}
    \end{minipage} 
\end{table}
\vspace{-4mm}
\section{Conclusions}

We propose to apply Few-Shot Learning to address the classification for rare skin diseases. We find that existing FSL methods do not perform significantly better than the baseline methods. Through careful analysis, we believe the problem should be largely attributed to the incompatibility between the episode training mechanism and cross entropy loss. Therefore, we propose a novel QR loss for FSL to make fully use of the information across samples and also allow the model to learn to extract information of the support samples guided by the training objective. With the proposed QR loss, the state-of-the-art FSL methods perform consistently better than methods training with the conventional CE loss. Our work demonstrates the promise of diagnosing rare skin diseases with one or a few labeled samples. In the future, we will investigate extensions to other medical classification problems or even natural image classification.  

\bibliographystyle{splncs04}
\bibliography{adasvm}
\end{document}